\documentclass[review,12pt]{elsarticle}
\usepackage{amsmath}
\usepackage{graphicx}
\usepackage{lineno,hyperref}
\modulolinenumbers[1]

\hypersetup{
    colorlinks=true,
    linkcolor=blue,
    filecolor=magenta,      
    urlcolor=blue,
    citecolor=blue,
}

\begin{document}

\begin{frontmatter}

\title{Strange behavior of rapidity dependent strangeness enhancement of particles containing and not containing leading quarks}

\author[mymainaddress,mysecondaryaddress]{Kalyan Dey\fnref{myfootnote}}
\author[mymainaddress]{B. Bhattacharjee \corref{mycorrespondingauthor}}
\cortext[mycorrespondingauthor]{Corresponding author}
\ead{buddhadeb.bhattacharjee@cern.ch}


\fntext[myfootnote]{k.dey@gsi.de}


\address[mymainaddress]{Nuclear and Radiation Physics Research Laboratory\\ Department of Physics, Gauhati University Guwahati, Assam 781014, India}
\address[mysecondaryaddress]{Department of Physics, Bodoland University, Rangalikhata, Deborgaon\\
Kokrajhar, Assam 783370, India}

\begin{abstract}
Rapidity dependent strangeness enhancement factors for the identified particles have been studied with the help of a string based hadronic transport model UrQMD-3.3 (Ultra-relativistic Quantum Molecular Dynamics) at FAIR energies. A strong rapidity dependent strangeness enhancement could be observed with our generated data for $Au + Au$ collisions at the beam energy of 30\textit{A} GeV. The strangeness enhancement is found to be maximum at mid-rapidity for the particles containing leading quarks while for particles consisting of produced quarks only, the situation is seen to be otherwise. Such rapidity dependent strangeness enhancement could be traced back to the dependence of rapidity width on centrality or otherwise on the distribution of net-baryon density. 
\end{abstract}


\end{frontmatter}


\section{Introduction}	
In heavy-ion collisions, the pattern of variation of net-baryon density ($\mu_B$) in rapidity space is found to vary with  beam energy. For example, at SIS18/AGS energies, the variation of net-baryon density with rapidity is found to be of Gaussian shape with its peak at mid-rapidity \cite{muB1}, whereas at top SPS and RHIC energies, such variation of $\mu_B$ shows bimodality with a minimum at mid-rapidity \cite{muB2, muB3, muB4, muB5, muB6, muB7}. At LHC energies, the net-baryon density is found to be close to zero at mid-rapidity \cite{LHC}. It is therefore easily comprehensible that the rapidity distribution of a particle, the production of which is sensitive to net-baryon density, or otherwise, the particles containing leading quarks (such as $k^+$, $\Lambda$, $\Xi^-$), might tend to follow the net-baryon density distribution. The rapidity dependence of other particles whose none of the constituents is $u$ or $d$ quark might not exhibit any such dependencies on net-baryon density.  Our earlier work \cite{dey} on the variation of rapidity width of various produced particles on beam rapidity, with special reference to $\Lambda$ and $\overline{\Lambda}$, has vindicated such prediction. It may, therefore, be not completely out of context to believe that a number of other kinematic and dynamical properties of heavy-ion collision might get coupled with this net-baryon density distribution effect.

Strange particles are produced only at the time of collisions and thus expected to carry important information of collision dynamics. Strangeness enhancement is considered to be one of the traditional signatures \cite{strange1, strange2} of formation of Quark Gluon Plasma (QGP). Due to the limitation of the detector acceptance, the past and ongoing heavy-ion experiments could measure the strangeness enhancement at mid-rapidity only. All such measurements assume a global conservation of strangeness. However, Steinheimer \textit{et al.} \cite{stein1} from UrQMD calculation predicted that strangeness is not uniformly distributed over rapidity space leading to a local violation of strangeness conservation. Thus, the study of rapidity dependent strangeness enhancement is of considerable significance.

Considering the fact that with the large acceptance detectors of the upcoming FAIR-CBM experiment \cite{cbm1, cbm2, cbm3, cbm4},  measurement of the rapidity dependent strangeness enhancement factor could be a possibility, strangeness enhancement factor at different rapidity bin has been estimated for various identified particles produced in $Au+Au$ collisions at 30\textit{A} GeV using a string based hadronic transport model UrQMD-3.3 (Ultra -relativistic Quantum Molecular Dynamics). 

\section{The UrQMD model}

Ultra-relativistic Quantum Molecular Dynamics (UrQMD) \cite{UrQMD1, UrQMD2} is a microscopic transport model based on a phase space description of $p+p$, $p + A$ and $A + A$ collisions that remains successful in describing the observables of heavy-ion collisions over a wide range of beam energies, that is, from $E_{lab}$ = 100 AMeV to $\sqrt{s}$ = 200 GeV \cite{U1,U2,U3,U4}. At low and intermediate energies ($\sqrt{s} <$  5 GeV), it describes the phenomenology of heavy-ion collisions in terms of interactions between known hadrons and their resonances. Fifty-five baryon species up to an invariant mass of 2.25 GeV and 32 meson species up to 2 GeV have been included
in this model. At higher energies ($\sqrt{s} >$ 5 GeV), the excitation of color strings and their subsequent fragmentation into hadrons are taken into account \cite{UrQMD1, UrQMD2, UrQMD3}.
The string models \cite{lund1} are found to be very successful in describing various dynamical features of high-energy heavy-ion collisions. One of the main ingredients of the string models is the \emph{string fragmentation function}. The fragmentation function in string models generally determines the kinematics of the produced particles. The fragmentation function $\mathrm{f(x)}$ represents the probability distribution for hadrons to acquire the longitudinal momentum fraction x from the fragmenting string \cite{UrQMD2}. Over the year, different types of fragmentation functions were proposed e.g. Lund-fragmentation \cite{lund1}, Field-Feynman fragmentation functions \cite{field} etc. In the default setting of the UrQMD model, Field-Feynman \cite{field} fragmentation scheme is used for the produced particles (see Eqn. \ref{eq5_1}) whereas for the leading nucleons, a different kind of fragmentation scheme is implemented (see Eqn. \ref{eq5_2}). The values of the free parameters of the fragmentation function have been optimized in accordance with the experimental data. For the sake of completeness, the fragmentation functions used in the default version of UrQMD model have also been plotted as a function of longitudinal momentum fraction $x$ as shown in Fig. \ref{SF}. The functional form of the fragmentation functions used in UrQMD model are given below,
\begin{align}
f(x)_{prod} &=  (1-x)^2, \label{eq5_1} \\ 
f(x)_{lead} &=  \exp \left[-\frac{(x-B)^2}{2A^2} \right], \label{eq5_2}
\end{align}
where A and B are free parameters, A = 0.275 and B = 0.42. The parameters A and B respectively represent the standard deviation and mean of the distribution as shown in Eqn. \ref{eq5_2}. $f(x)_{prod}$ and $f(x)_{lead}$ are the fragmentation function of produced particles and leading nucleons respectively.

\begin{figure*}[h] 
\centering
\includegraphics*[width=0.9\textwidth]{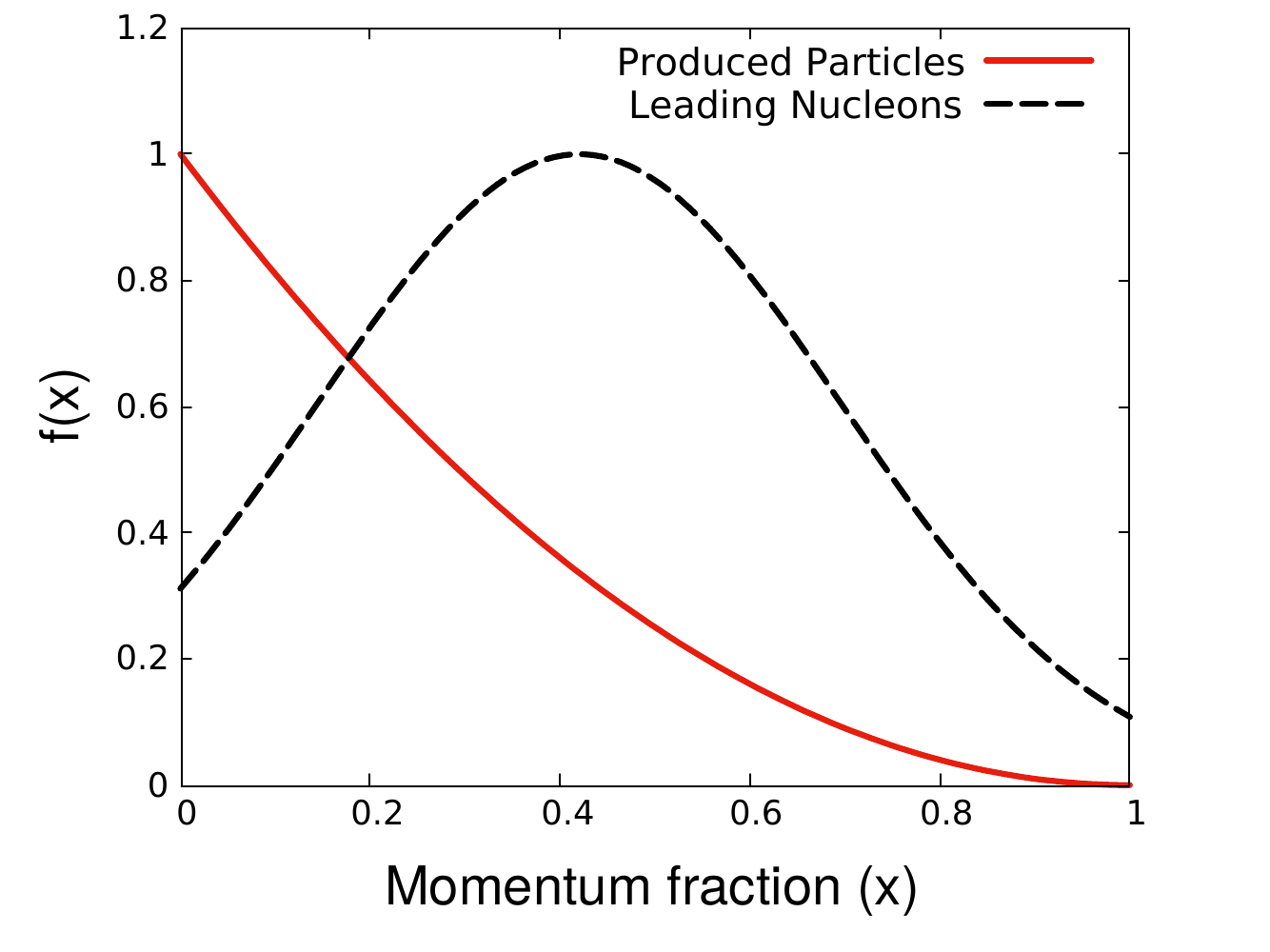} 
\caption{String fragmentation functions used in the default version of UrQMD model \cite{UrQMD1}.} \label{SF}
\end{figure*}

\textbf{Strange particle production in UrQMD:} As far as the production of strange hadrons is concerned, in heavy-ion collisions, they may either be produced in the very early stage in the initial collisions among the incoming nucleons or at the later stage through secondary collisions among the produced particles \cite{Bass1}. In low energy domain, that is, close to threshold energy, strange particles can either be produced in NN reaction channel directly,  also known as \emph{associated strangeness production}, e.g., $p + p \rightarrow p + \Lambda + K^+$ or in two steps reactions such as $p + p \rightarrow N + N^*_{1710}$ and finally $N^*_{1710} \rightarrow Y + K^+$. As already discussed, at higher incident beam energies, strange particle production, in general, is dominated by string excitation and fragmentation \cite{Bass1}. It is to be noted here that other than string excitation and fragmentation, the multi-strange hyperons like $\Omega^-$(sss) can also be produced via strangeness exchange reaction channels like $\Xi K^− \rightarrow \Omega \pi$.

\section{Strangeness enhancement}
Strangeness enhancement factor $E_S$ is quantified by measuring the ratio of the yield of strange particles in $A+A$ collisions and respective yield in $p+p$ collisions, where both the numerator and denominator are normalized by the average number of participating nucleons $\langle N_{part} \rangle$, calculated using Glauber model \cite{Glauber}. The conventional definition of strangeness enhancement factor is  \cite{def1, def2},

\begin{equation}
E_{S} = \frac{(Yield)_{AA}}{\langle N_{part} \rangle}/ \frac{(Yield)_{pp}}{2}.   
\end{equation}

In this report, using  reference \cite{alt_def}, an alternative definition of strangeness enhancement factor $E_S$ is used where $E_S$ is defined as,

\begin{equation}
E_{S} = \left[\frac{(Yield)_{AA}}{\langle N_{\pi^-}\rangle}\right] _{central}/ \left[\frac{(Yield)_{AA}}{\langle N_{\pi^-}\rangle}\right] _{peripheral},   
\label{r}
\end{equation}

The reason for taking the average number of produced pions $\langle N_{\pi^-} \rangle$, instead of $\langle N_{part}\rangle$, as a measure of centrality variable is described elsewhere \cite{non_linear, dae_ref}.

\section{Results and discussion}
\begin{figure*}[h]
\includegraphics*[width=\textwidth]{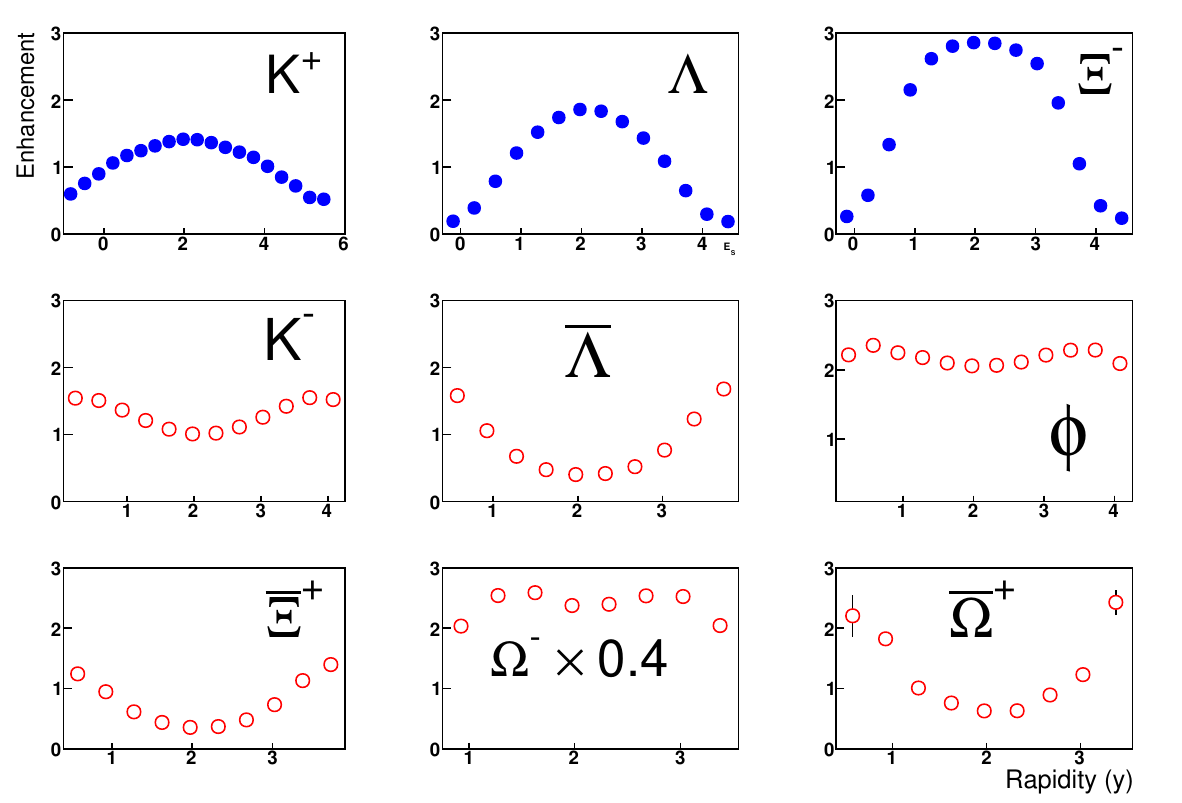}
\caption{\label{fig1} Strangeness enhancement factor as a function of rapidity for particles containing at least one leading quark (filled circle) and particles containing only produced quarks (open circle) for $Au+Au$ collisions at 30\textit{A} GeV. The errors are small and are within the symbol size.}
\end{figure*}

A total of 93 million minimum biased UrQMD events in default mode has been used for the present analysis.
The strangeness enhancement factor ($E_S$) has been estimated and plotted as a function of rapidity for various identified strange particles for $Au+Au$ collisions at 30\textit{A} GeV and is shown in Fig. \ref{fig1}. The impact parameter windows chosen for central and peripheral collisions are 0-4.1 fm and 9-13 fm respectively. We choose $Au+Au$ collisions at 30\textit{A} GeV for the present study as the baryon density is reported to be maximum at this energy \cite{baryon_density_cbm}. It is interesting to see from the figure that $E_S$ depends strongly on the rapidity and this dependence follow two distinctive patterns. While the enhancement factor at mid-rapidity is found to be maximum for the particles containing leading quarks (filled circle), the same is observed to be minimum at mid-rapidity for the particles containing produced quarks only (open circle). However, even though a rise and fall pattern is also visible for $\Omega^-$, consisting of three produced quarks only (sss), the existence of a slight dip at mid-rapidity can be clearly visible which is consistent with the general trend and would be discussed in detail later.

\begin{figure*} [h]
\includegraphics*[width=\textwidth]{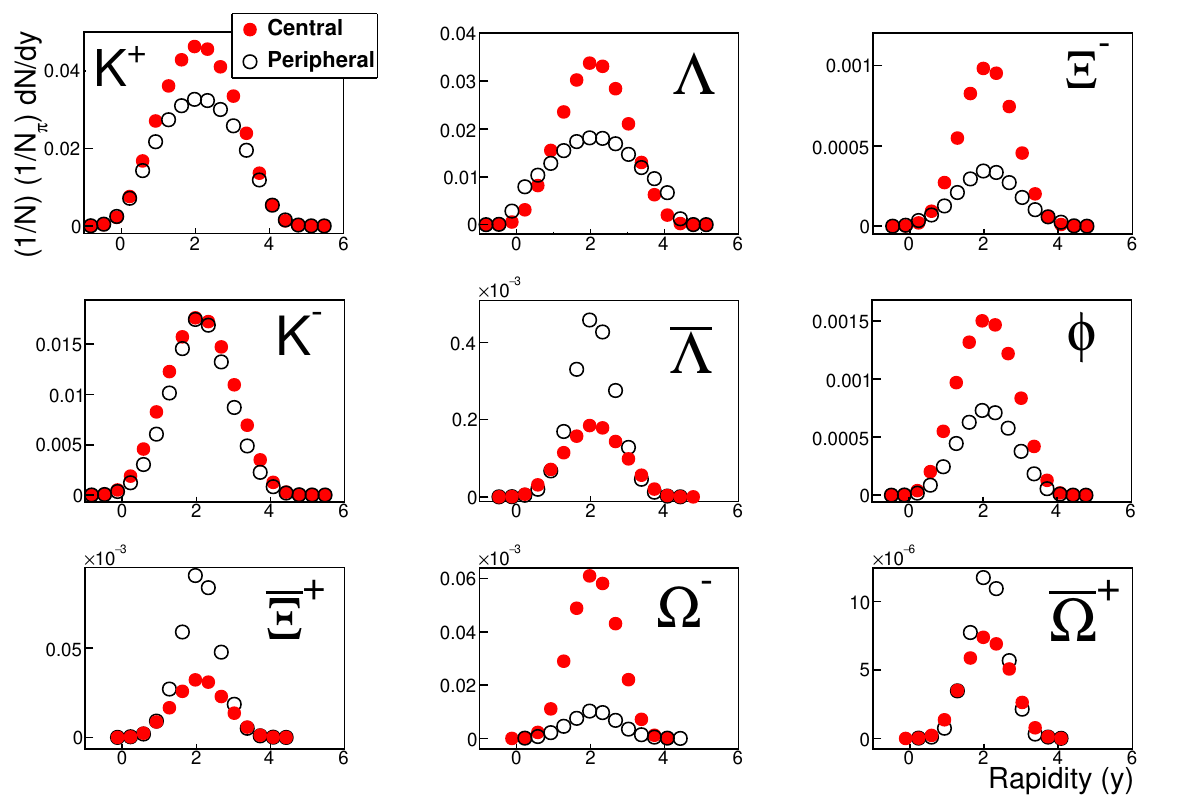}
\caption{Normalized rapidity distribution of particles containing leading and produced quarks for central as well as peripheral $Au+Au$ collisions at 30\textit{A} GeV. The errors are small and are within the symbol size.}
\label{fig2}
\end{figure*}

\begin{figure*} [tb]
\includegraphics*[width=\textwidth]{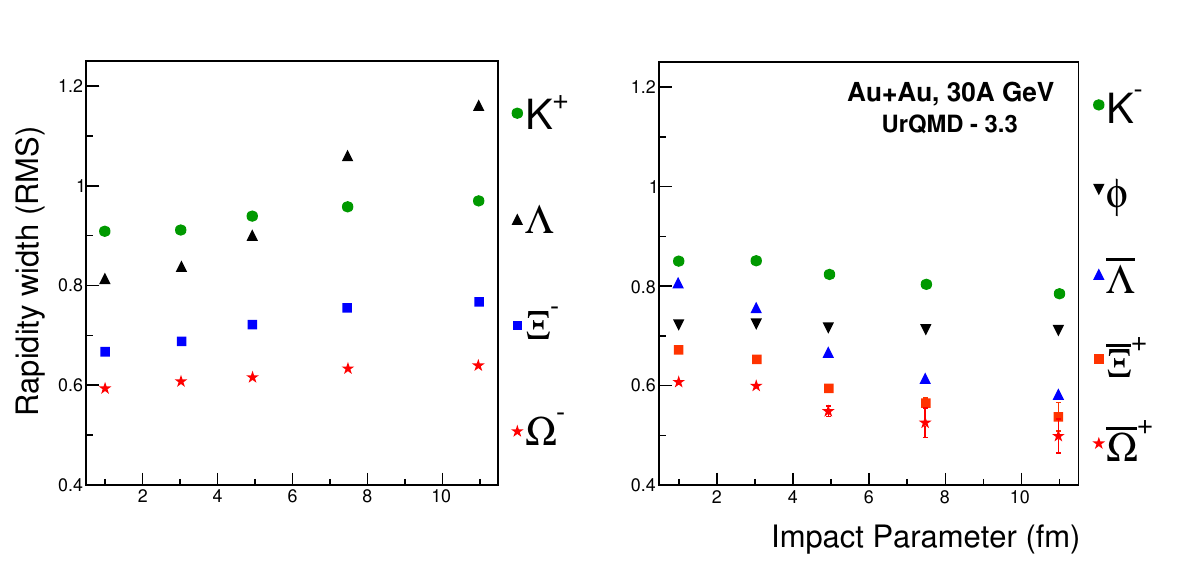}
\caption{\label{fig3} Rapidity width as a function of impact parameter for identified particles for $Au+Au$ collisions at 30\textit{A} GeV. The rapidity width or RMS has been calculated by parameterizing the rapidity distributions with a double Gaussian. The details of the parameterization is provided in the ref. \cite{dey}.}
\end{figure*}

\begin{figure*} [tb]
\begin{center}
\includegraphics[width=\textwidth]{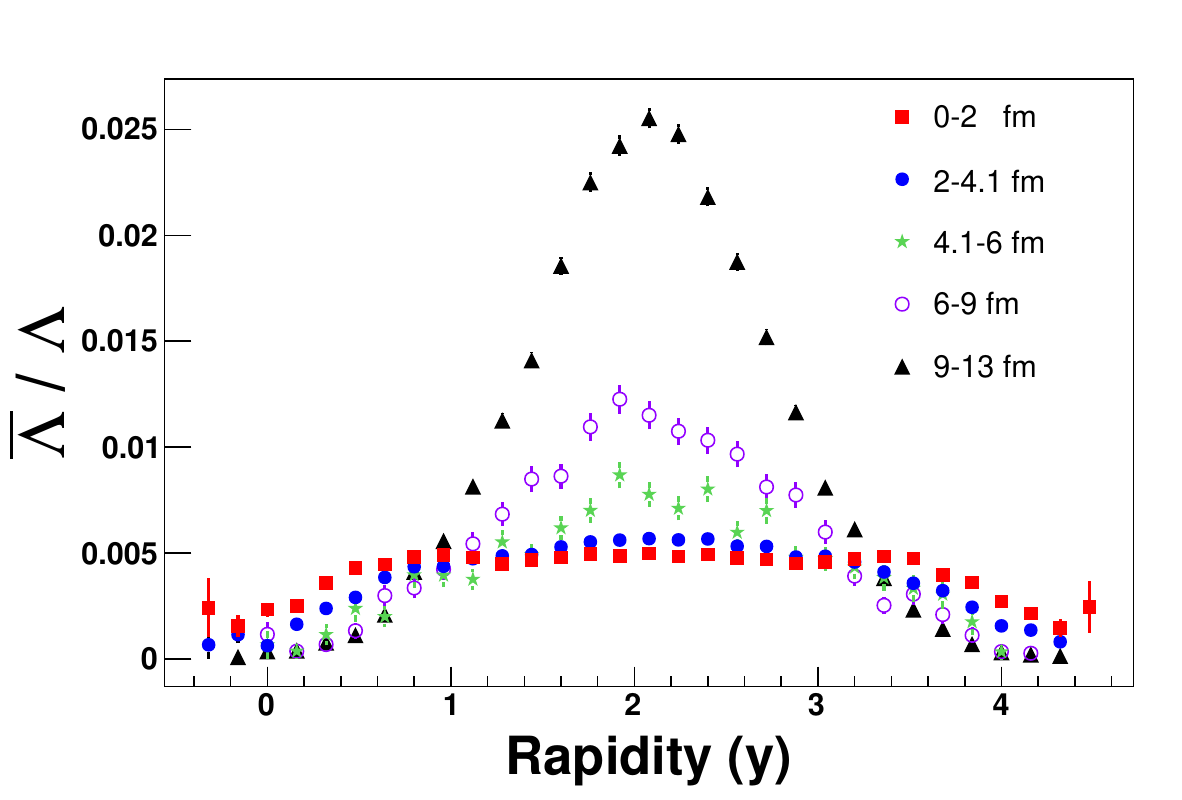}
\caption{$\bar{\Lambda}/\Lambda$ as a function of rapidity for different centrality interval in $Au+Au$ collisions at $E_{lab} = 30A$ GeV.}
\label{fig4}
\end{center}
\end{figure*}

\begin{figure} [h]
\begin{center}
\includegraphics[width=0.8\textwidth]{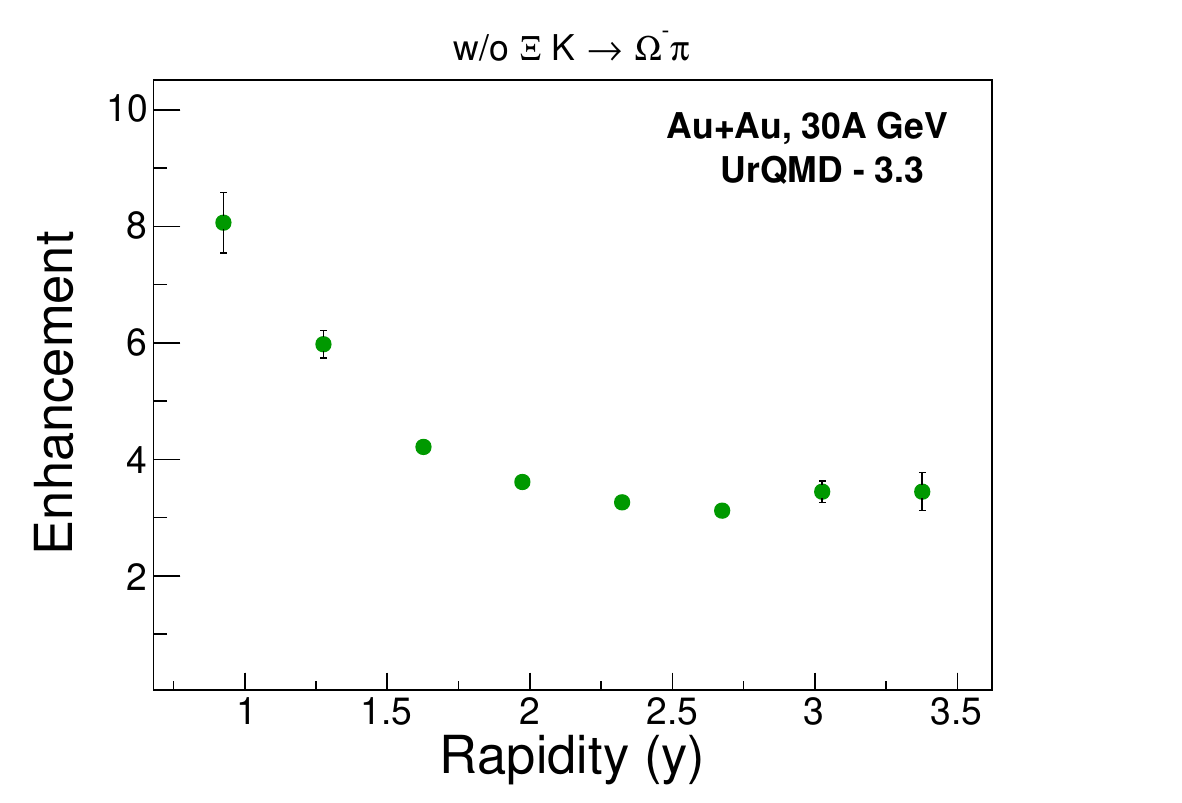}
\caption{\label{fig5} Strangeness enhancement factor for $\Omega^-$ as a function of rapidity for $Au+Au$ collisions at 30\textit{A} GeV without the $\Xi K \rightarrow \Omega^- \pi$.}
\end{center}
\end{figure}

\begin{figure*} [h]
\begin{center}
\includegraphics[width=\textwidth]{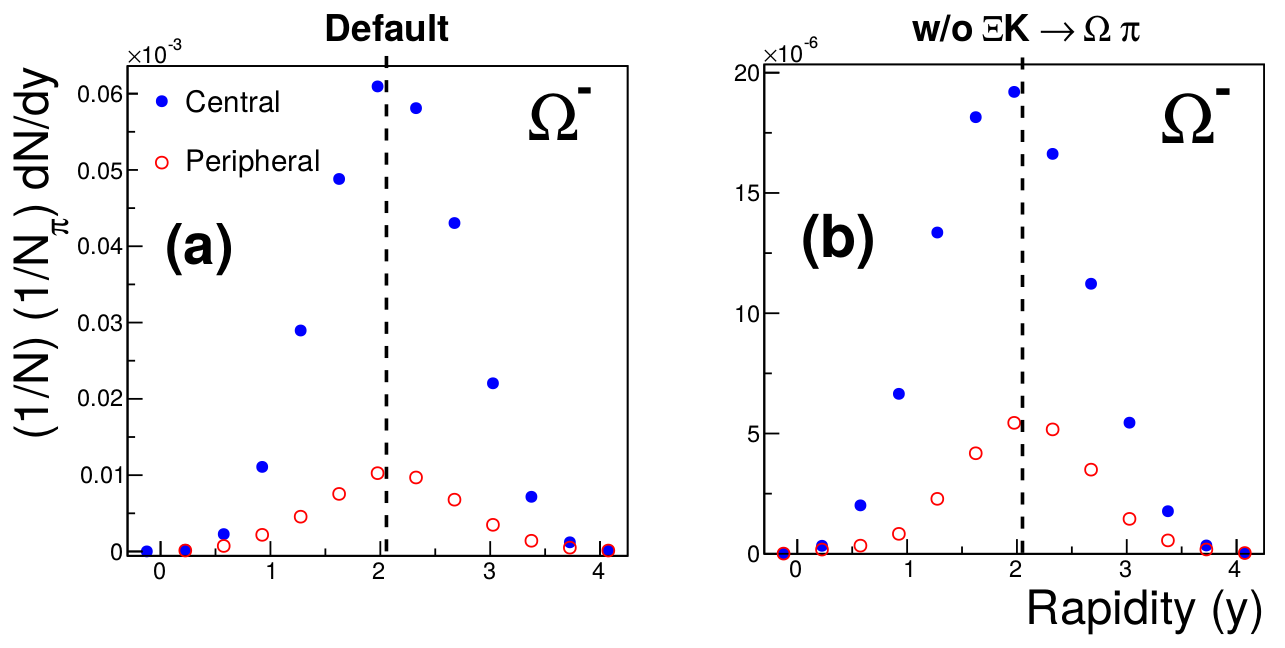}
\caption{\label{fig6} Normalized rapidity distribution of $\Omega^-$ baryon in central and peripheral collisions (a) for default UrQMD and (b) by switching off the $\Xi K \rightarrow \Omega\pi$ channel in $Au + Au$ collisions at 30\textit{A} GeV. The errors are small and are within the symbol size.}
\end{center}
\end{figure*}

To understand the underlying dynamics of such rapidity dependent strangeness enhancement, the normalized rapidity distributions for both central and peripheral collisions (Fig. \ref{fig2}) and the width of the rapidity distribution as a function of centrality (Fig. \ref{fig3}) are plotted for various produced particles. It is readily evident from Fig. \ref{fig2} and Fig. \ref{fig3} that the different patterns of variation of rapidity dependent strangeness enhancement factor for particles containing and not containing leading quarks lie on the dependence of rapidity distribution or otherwise the variation of the width of the rapidity distribution on centrality. The width of the rapidity distribution increases as we go from central to the peripheral collisions for the particles containing leading quarks, a feature that might be attributed to the variation of net-baryon density with impact parameter at the studied energy. In Fig. \ref{fig4}, we, therefore, plot  $\bar{\Lambda}$ to $\Lambda$ ratio against rapidity for different centrality for the studied $Au+Au$ collisions at 30\textit{A} GeV. It could be readily seen from this figure that as we go from central to peripheral collision, the width of the rapidity distribution of $\bar{\Lambda}$ to $\Lambda$ ratio decreases. Thus, with increasing impact parameter, as expected, the net-baryon number tends to populate the extreme rapidity spaces due to small overlap of the colliding partners. This increase in the population of baryons over anti-baryons at larger rapidities causes the broadening of the width of the rapidity distribution of the particles containing leading quarks (since their production is dependent on the distribution of $u$ and $d$ quarks). This may otherwise mean that at the studied energy, a large fraction of such particles are not pair produced. Further, from Fig. \ref{fig3} (right panel) it is also seen that for all but $\Omega^-$ particles, consisting of produced quarks only, the rapidity widths decrease with decreasing centrality (or increasing impact parameter). Such decrease in the width of the rapidity distribution of these particles could be due to the decrease in the size of the central fireball with the increase of impact parameter. It is to be noted here that, the observed dependence of width of the rapidity distribution with centrality has also been reported for kaons and lambdas by the NA49 collaboration for 40\textit{A} GeV Pb+Pb collisions \cite{sys1, sys2}. Moreover, for $\Omega^-$ baryon, whose all the constituent quarks are produced quarks only (sss), its production in UrQMD-3.3, in addition to string fragmentation, is also influenced by $\Xi^-$ and $\Xi^0$ via $\Xi^- + K^- \rightarrow \Omega^- + \pi^-$, $\Xi^- + \bar{K^0} \rightarrow \Omega^- + \pi^0$, $\Xi^0 + K^- \rightarrow \Omega^- + \pi^0$, and $\Xi^0 + \bar{K^0} \rightarrow \Omega^- + \pi^+$. Both $\Xi^-$ (dss) and $\Xi^0$ (uss) contain a leading quark as one of their constituents. It is because of this influence of $\Xi^-$ ($\Xi^0$), $\Omega^-$ is found to exhibit a different pattern other than its species members (having no leading quarks). To prevent the production of $\Omega^-$ from $\Xi K$ interactions, another 60 millions of minimum biased events were generated by switching off the reaction channels viz. $\Xi^- + K^- \rightarrow \Omega^- + \pi^-$, $\Xi^- + \bar{K}^0 \rightarrow \Omega^- + \pi^0$, $\Xi^0 + K^- \rightarrow \Omega^- + \pi^0$, $\Xi^0 + \bar{K}^0 \rightarrow \Omega^- + \pi^+$, $\bar{\Xi}^+ + K^+ \rightarrow \bar{\Omega}^+ + \pi^+$, $\bar{\Xi}^+ + K^0 \rightarrow \bar{\Omega}^+ + \pi^0$, $\bar{\Xi}^0 + K^+ \rightarrow \bar{\Omega}^+ + \pi^0$, $\bar{\Xi}^0 + K^0 \rightarrow \bar{\Omega}^+ + \pi^-$ in the UrQMD event generator and the resulting rapidity dependent enhancement factor of $\Omega^-$ is re-plotted in Fig. \ref{fig5}. It can be readily seen from this figure that the generic rise and fall pattern of $E_S$, as seen for leading particles, is now completely missing. Even though the plot is not exactly symmetric, which could be due to the fact that UrQMD does not symmetrize when it does kinematics in absence of the aforementioned reaction channels, as is evident from the Fig. \ref{fig6}, the presence of a clear minimum around mid-rapidity can surely be not ruled out.

\begin{figure} [h]
\centering
\includegraphics[width=\textwidth]{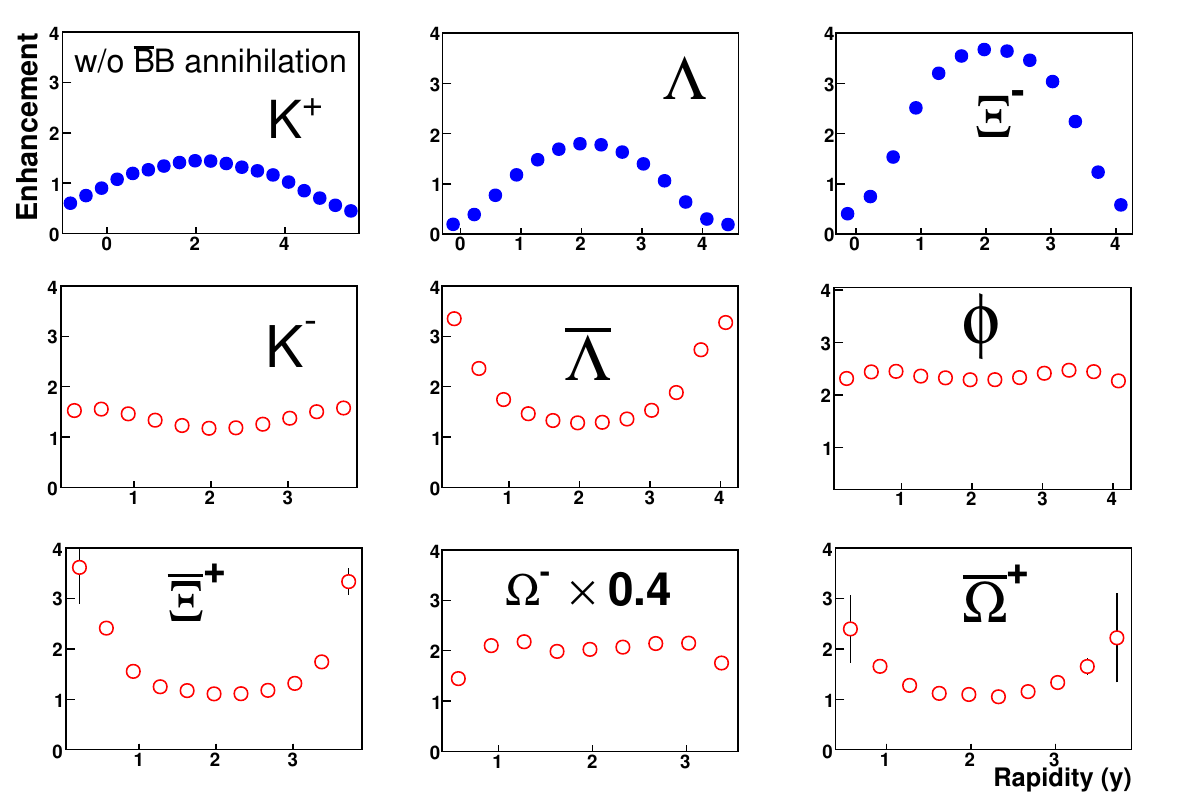}
\caption{Strangeness enhancement factor as a function of rapidity for particles containing at least one leading quark (filled circle) and particles containing only produced quarks (open circle) for Au + Au collisions at 30A GeV by switching off B$\bar{\mathrm{B}}$ anihilation.} \label{BB-an}
\end{figure}

As from Fig. \ref{fig4}, it is seen that anti-particle to particle ratio is much less than unity, annihilation processes are expected to be negligible in leading quark hadrons while they could be dominant for non-leading quark hadrons. To ascertain the role of annihilation process, if any, on strangeness enhancement factor, another 20 million UrQMD events at 30\textit{A} GeV are generated by turning off the annihilation process and rapidity dependent enhancement factors $E_S$ for various studied hadrons have been re-plotted in Fig. \ref{BB-an}. From this figure, it could be readily seen that the stopping of annihilation processes indeed has no significant effect on $E_S$ of leading quark hadrons. However, for non-leading quark hadrons, change in $E_S$ is much more significant. Even though the general shape of the rapidity dependent $E_S$ for non-leading hadrons remains the same i.e., minimum at mid-rapidity, with the stopping of annihilation process, the enhancement factors at mid-rapidity have been increased considerably changing the suppression ($E_S<1$) to enhancement ($E_S>1$).

Further, in order to investigate the influence of the string fragmentation schemes on the observed strangeness enhancement pattern, another 45 million events were generated by implementing the Lund-fragmentation function as string fragmentation scheme for the produced particles in the UrQMD model. The Lund string fragmentation function is given by \cite{def2},
\begin{align}
f(x) \propto \frac{1}{x}  (1-x)^a \exp \left(-\frac{bm_T^2}{x}  \right),
\end{align}

where $a$ and $b$ are the free parameters to be fixed by the experimental data. $m_T$ represents the transverse mass of the produced hadron. The average squared transverse momentum $\langle p_T^2\rangle$ of produced particle is proportional to the string tension $\kappa$. The two parameters $a$ and $b$ are approximately related to the string tension by the following relation \cite{lund1},
\begin{align}
\kappa \propto \left[ b(2+a) \right]^{-1}.
\end{align}

\begin{figure} [tb]
\centering
\includegraphics[width=\textwidth]{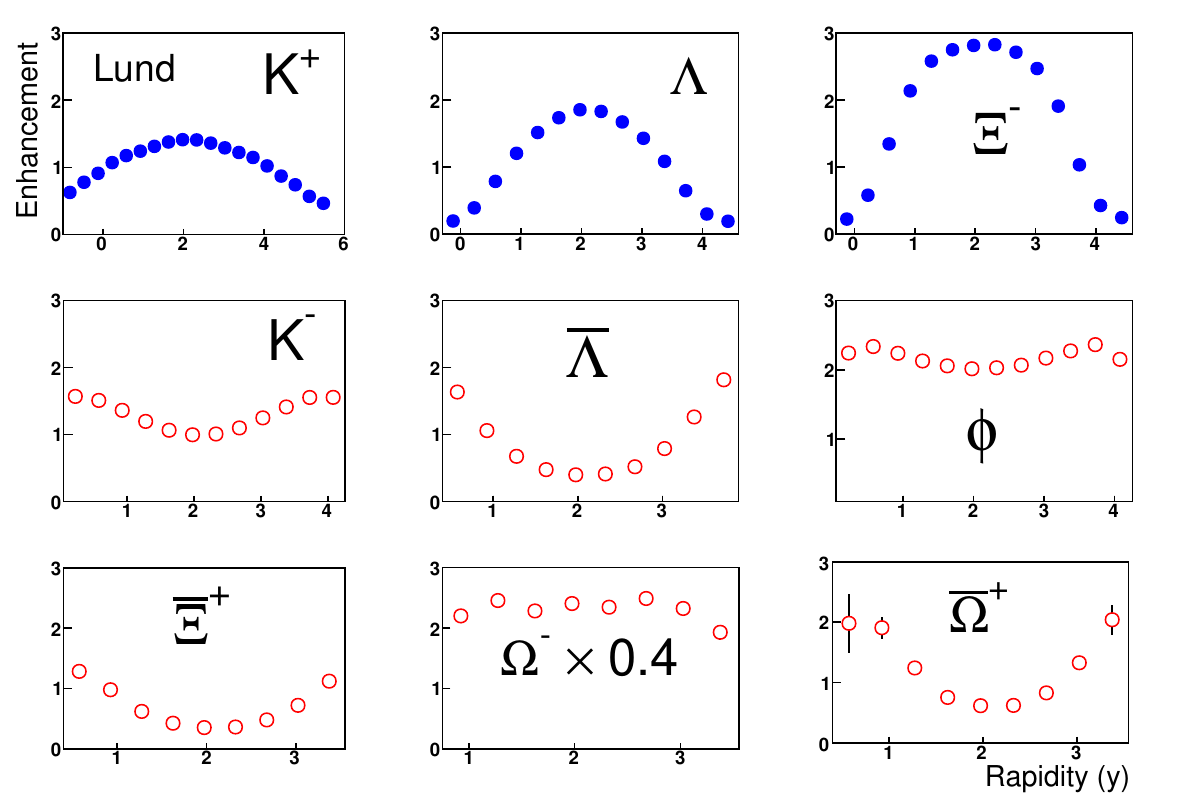}
\caption{Strangeness enhancement factor as a function of rapidity for particles containing at least one leading quark (filled circle) and particles containing only produced quarks (open circle) for Au + Au collisions at 30A GeV using Lund fragmentation function.} \label{LUND1}
\end{figure}

Fig. \ref{LUND1} illustrates the rapidity dependent strangeness enhancement for the produced particles using Lund fragmentation model as the string fragmentation scheme. It is clearly seen from the figure that, similar to that of Field-Feynman fragmentation scheme (blue solid line in Fig. \ref{SF}), the LUND model also predicted two distinctive patterns of rapidity dependent strangeness enhancement for the particles containing and not containing leading quarks.

\section{Summary}
Strangeness enhancement factors of various produced particles containing and not containing leading quark(s) as one of the constituents are studied at various rapidity bins for $Au+Au$ collisions at 30\textit{A} GeV. It is interesting to observe from this investigation that the patterns of rapidity dependent strangeness enhancement factors depend  on the quark content of hadrons. At mid-rapidity, the strangeness enhancement factor is found to be maximum for the particles containing leading quark(s) while it shows a minimum at mid-rapidity for the particles containing produced quarks only. Such two distinctive patterns of variation of strangeness enhancement factor with rapidity have been attributed to the fact that the variation of rapidity width of the produced particles follows two different patterns with the centrality of collision for particles containing or not containing leading quarks which in turn, is dependent on the variation of net-baryon density distribution with collision centrality. Moreover, even though the annihilation process is found to have little effect on $E_S$ of leading quark hadrons, it considerably changes the $E_S$ of non-leading quark hadrons at all rapidity space. Further, the observed rapidity dependent strangeness enhancement pattern is found to be independent of the string fragmentation schemes of UrQMD considered for particle production.

\section*{Acknowledgments}
The authors would like to offer their sincere thanks to Prof. S.A. Bass, and Prof. Volker Friese for some meaningful discussions and valuable  suggestions. The authors are also thankful to Gunnar Gr\"af for helping them in incorporating some relevant changes in UrQMD code. The help received from N. Hussain and S. Gope are appreciated for generating a part of UrQMD events, using the HPCC facility of Nuclear and Radiation Physics Research Laboratory, Department of Physics, Gauhati University, Guwahati, created with the financial assistance of DTS, Govt. of India (vide project No. SR/MF/PS-012009). The authors also appreciate the support received from GSI Helmholtzzentrum f\"ur Schwerionenforschung, Darmstadt, Germany for providing the computing resources for generating UrQMD events (default) used in this analysis.




\end{document}